\documentclass[3p]{elsarticle}
\usepackage[T1]{fontenc}
\usepackage{amsfonts,amsmath,amssymb,bm,slashed}
\usepackage{graphicx,xcolor,microtype}
\usepackage{float}
\usepackage{multicol}
\raggedcolumns
\usepackage[hidelinks]{hyperref}
\journal{Physics Letters B}
\allowdisplaybreaks[1]
\newcommand{\dd}{\mathrm{d}}
\newcommand{\ii}{\mathrm{i}}
\newcommand{\tr}{\operatorname{tr}}
\newcommand{\im}{\operatorname{Im}}

\newcommand{\bq}{\bm q}
\newcommand{\bk}{\bm k}
\newcommand{\br}{\bm r}
\newcommand{\bp}{\bm p}
\newcommand{\intq}{\int\!\frac{\dd^3\bq}{(2\pi)^3}}
\newcommand{\intk}{\int\!\frac{\dd^3\bk}{(2\pi)^3}}
\newcommand{\mq}{m_{D,q}}
\newcommand{\nf}{n_F}
\newcommand{\M}{\mathcal M}

\newcommand{\ket}[1]{\lvert #1\rangle}
\newcommand{\bra}[1]{\langle #1\rvert}
\newcommand{\avg}[1]{\langle #1\rangle} 

\begin{document}

\begin{frontmatter}

\title{The imaginary heavy quark--antiquark potential in the quark--gluon plasma revisited}

\author[a]{Min He}

\address[a]{Department of Applied Physics, Nanjing University of Science and Technology, Nanjing 210094, China}

\begin{abstract}
In a weakly coupled Quark–-Gluon Plasma, Landau damping gives the static heavy quark–-antiquark ($Q\bar Q$) potential an imaginary part that attenuates the pair amplitude. We present a new derivation of the  $Q\bar Q$ imaginary potential from the perspective of unitarity in terms of the in-medium generalized optical theorem expressed in the joint heavy-pair and thermal bath Hilbert space. The two Landau-cut assignments of the non-forward exchange amplitude yield interference between scatterings of $Q$ and $\bar Q$ from bath particles. The bath trace supplies appropriate thermal weights for these interference terms. Combining the color-projected interference terms with the single heavy-source self-energies recovers the leading-order static imaginary potential in both color-singlet and -octet channels. We therefore attributes the origin of the imaginary $Q\bar Q$ potential essentially and explicitly to an interference mechanism as facilitated by thermal bath particles. 

\end{abstract}


\end{frontmatter}
\raggedbottom
% Keep column flow continuous; full-width panels temporarily end multicols.
\begin{multicols}{2}

%%%%%%%%%%%%%%%%%%%%%%%%%%%%%%%%%%%%%%%%%%%%%%%%%%%%%%%%
\section{Introduction}
\label{sec_intro}
%%%%%%%%%%%%%%%%%%%%%%%%%%%%%%%%%%%%%%%%%%%%%%%%%%%%%%%%
The large quark mass and small relative velocity of a bound heavy quark ($Q$) and antiquark ($\bar Q$) allow their relative motion to be described, at leading nonrelativistic order, by a Schr\"odinger equation with a potential. In vacuum, the Cornell potential $V(r)=-\kappa/r+\sigma r$ combines a short-distance Coulomb attraction with a linear confining term of string tension $\sigma$~\cite{Eichten1978}. In a Quark--Gluon Plasma (QGP), mobile color charges screen the interaction. Matsui and Satz~\cite{Matsui1986} argued that static screening of the color-singlet $Q$-$\bar Q$ potential, parameterized as $V(r,m_D)=-\kappa e^{-m_Dr}/r+\sigma(1-e^{-m_Dr})/m_D$ with a temperature-dependent screening mass $m_D$, would prevent the formation of heavy quarkonia when the screening length becomes shorter than the bound-state radius. More modern perspective is, when placed in the QGP, the $Q\bar Q$ pair exchanges energy and color with the medium and thus acts as an open quantum system~\cite{Brambilla:2016wgg,Blaizot:2017ypk,Yao:2021lus,Akamatsu2022}. Tracing out the medium degrees of freedom therefore generally produces non-unitary evolution for the $Q\bar Q$ pair. When a local potential description applies, the reduced pair dynamics is thus encoded in a complex potential, with the real part governing the coherent interaction and its imaginary part accounting for the attenuation of the pair amplitude.

Laine {\it et al.}~\cite{Laine2007} first derived the complex static \(Q\bar Q\) potential in thermal perturbative QCD, and Beraudo {\it et al.}~\cite{Beraudo2008} obtained its analogue for two static charges in a hot QED plasma. Both derivations identify the potential through the real-time evolution of a greater heavy-pair correlator \(G(t,r)\). After the trivial static propagation factors are removed, its logarithmic derivative defines $V(t,r)=\ii\partial_t\ln G(t,r)$, with $G(t,r)\sim e^{-\ii V(r)t}$ at late times. Laine {\it et al.}~\cite{Laine2007} started from a point-split heavy-current correlator that reduces to a Wilson loop in the infinite-mass limit. They evaluated the loop in imaginary time using hard-thermal-loop (HTL) perturbation theory and analytically continued it to real time. In the long-time limit, they found a Debye-screened real part and an imaginary part that gives quarkonium a thermal width. The latter arises from Landau damping of the exchanged spacelike gluon, through which the soft gauge field transfers energy to hard plasma constituents. In the QED derivation, Beraudo {\it et al.}~\cite{Beraudo2008} performed a Gaussian HTL average over the background fields, expressing the heavy-pair propagator as an exponential involving the time-ordered photon propagator. Its long-time exponent yields the full complex potential and relates the damping to collisions with light plasma constituents. Lattice QCD studies have since investigated the complex static potential through Euclidean Wilson loops; the extraction of its imaginary part is particularly sensitive to how well the spectral widths can be resolved~\cite{Burnier2015,Bala2022,Bazavov2024,Ali:2025iux}. The static complex potential was also extracted from a thermodynamic $T$-matrix formalism, where the imaginary part of the potential was found to induce marked deviation of the real part from the free energy~\cite{Liu:2015ypa}.

In this work, we present a novel derivation of the in-medium static $Q\bar Q$ imaginary potential from the perspective of unitarity expressed in the combined heavy-pair and medium Hilbert space. We still work in the leading-order perturbation theory for a weakly coupled QGP as in the early derivations~\cite{Laine2007,Beraudo2008}. In our approach, through the thermal generalized optical theorem, we resolve the Landau cuts of the  $Q$-$\bar Q$ absorptive nonforward exchange amplitude. The resulting products of real elastic amplitudes are explicitly identified with  interference between scatterings of $Q$ and $\bar Q$ from thermal bath particles. This provides the methodological distinction from the two early correlator-based derivations: we construct the connected exchange kernel $\im\widetilde V_{\mathrm{ex}}$ explicitly from elementary scattering-amplitude interference rather than extract it from a long-time correlator. The bath trace determines Pauli blocking and Bose enhancement, while explicit color wave functions and the antiquark normal-ordering sign fix the interference factors. We then combine these exchange kernels with the independently calculated heavy source self-energies to recover the complete singlet and octet imaginary potentials. The construction thus attributes the origin of the imaginary $Q\bar Q$ potential essentially to an interference mechanism as assisted by the thermal bath. 

%%%%%%%%%%%%%%%%%%%%%%%%%%%%%%%%%%%%%%%%%%%%%%%%%%%%%%%%%%%%%%%%
\section{Unitarity and the generalized optical theorem}
\label{sec:unitarity}
%%%%%%%%%%%%%%%%%%%%%%%%%%%%%%%%%%%%%%%%%%%%%%%%%%%%%%%%%%%%%%%
With the scattering matrix $\hat S=1+\ii\hat T$, unitarity implies
\begin{equation}
	\hat T-\hat T^\dagger=\ii\hat T^\dagger\hat T,\qquad
	T_{fi}-T_{if}^*=\ii\sum_n T_{nf}^*T_{ni},
	\label{eq:vacuumOT}
\end{equation}
where $\ket{i}$ and $\ket{f}$ denote the initial and final state, respectively, and the summation over intermediate states $\ket{n}$ includes all discrete quantum numbers and the appropriate phase-space integrations. For a forward amplitude, Eq.~\eqref{eq:vacuumOT} reduces to the usual positive inclusive sum, $2\im T_{ii}=\sum_n|T_{ni}|^2$. For nonforward matrix elements, it is an interference relation and the right-hand side need not be positive term by term. When applying Eq.~(\ref{eq:vacuumOT}) to the case of $Q{\bar Q}$ scattering in vacuum, the light-quark polarization insertion in the $s$ channel has a pair-production cut for timelike momentum above threshold, $q^2\geq4m_q^2$, which is not of potential nature. On the other hand, the $t$ channel scattering driven by the exchange of a spacelike gluon has no absorptive part; consequently, the light-quark polarization insertion in this channel cannot generate a vacuum imaginary exchange potential.    

\begin{figure}[H]
	\centering
	\includegraphics[width=\columnwidth]{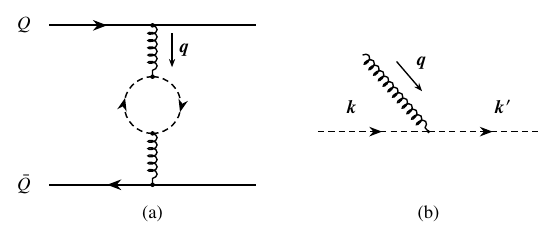}
	\caption{(a) Light-quark polarization insertion in the gluon exchanged between $Q$ and $\bar Q$ scattering in the medium. Solid lines denote the heavy quark and antiquark, dashed lines the light quark, and coils the gluon. (b) Absorption of the exchanged gluon by a bath quark underlying Landau damping, Eq.~\eqref{eq:landauprocess}, with $k'=k+q$. The bath-quark momenta are on shell, the incoming gluon is spacelike, and arrows indicate momentum flow.}
	\label{fig:landau-damping}
\end{figure}

However, in a medium, the gluon exchanged between $Q$ and $\bar Q$ still in the $t$ channel scattering (see Fig.~\ref{fig:landau-damping}(a)) can transfer energy and momentum to a bath quark already present in the initial state, as illustrated in Fig.~\ref{fig:landau-damping}(b),
\begin{equation}
	q(k)+g^*(q)\longrightarrow q(k'),\qquad k'=k+q,
	\label{eq:landauprocess}
\end{equation}
with both light-particle momenta on shell. For massless quarks, the condition $\omega=|\bk+\bq|-k$ has support in the spacelike region $|\omega|\leq q$. This is scattering from the bath, rather than the decay of a spacelike gluon into a newly created quark--antiquark pair. This process known as Landau damping~\cite{Laine2007} gives the exchanged gluon an absorptive self-energy and implies the emergence of an imaginary $Q\bar Q$ potential.

To incorporate this into the formalism of the generalized optical theorem Eq.~\eqref{eq:vacuumOT}, we consider a homogeneous, isotropic, color-neutral plasma in equilibrium at temperature $T$, with $N_f$ massless light flavors and vanishing chemical potentials. Let $\ket{i}$, $\ket{f}$, and $\ket{n}$ now denote the heavy-pair $Q\bar Q$ states, and $\ket{a}$ and $\ket{b}$ label states in a complete many-body basis of the medium. The medium equilibrium density operator is
\begin{equation}
\begin{aligned}
\rho =\sum_a p_a\ket{a}\bra{a},\qquad
 p_a=\frac{e^{-\beta E_a}}{Z},
\end{aligned}
\end{equation}
with $Z$ being the medium partition function, $E_a$ the medium energy of the state $\ket a$ and $\beta=T^{-1}$ the inverse temperature. 
Applying Eq.~\eqref{eq:vacuumOT} in the combined heavy-pair and bath Hilbert space gives
\begin{equation}
	T_{fa,ia}-T_{ia,fa}^*
	=\ii\sum_{n,b}T_{nb,fa}^*T_{nb,ia},
	\label{eq:TminusTstar}
\end{equation}
where the completeness relation $\sum \ket{n,b}\bra{n,b}=1$ has been inserted.

Defining the thermally averaged $Q\bar Q$ elastic amplitude
\begin{equation}
	\overline T_{fi}=\sum_a p_a T_{fa,ia},
	\label{eq:Tthermalaverage}
\end{equation}
we multiply both sides of Eq.~\eqref{eq:TminusTstar} by $p_a$ and sum over the initial bath states $\ket a$ to obtain
\begin{equation}
	\overline T_{fi}-\overline T_{if}^{\,*}
		=\ii\sum_{a,b,n}p_a\,T_{nb,fa}^*T_{nb,ia},
	\label{eq:thermalOT}
\end{equation}
which represents the in-medium version of the generalized optical theorem. In particular, the right-hand side is an average of products of amplitudes and cannot be replaced by a product of separately averaged amplitudes. The use of scattering notation in Eq.~\eqref{eq:thermalOT} is perturbative. It does not assume an independent unitary $S$ matrix for the reduced heavy-pair subsystem. Indeed, attenuation appears precisely because the evolution of that subsystem is not closed.

%%%%%%%%%%%%%%%%%%%%%%%%%%%%%%%%%%%%%%%%%%%%%%%%%%%%%%%%%%%%%%%%
\section{Derivation of the connected exchange imaginary $Q\bar Q$ potential}
\label{sec:derivation}
%%%%%%%%%%%%%%%%%%%%%%%%%%%%%%%%%%%%%%%%%%%%%%%%%%%%%%%%%%%%%%%%

%%%%%%%%%%%%%%%%%%%%%%%%%%%%%%%%%%%%%%%%%%%%%%%%%%%%%%%%%%%%%%%
\subsection{The effective imaginary $Q\bar Q$ potential}
\label{ssec:effective-potential}
%%%%%%%%%%%%%%%%%%%%%%%%%%%%%%%%%%%%%%%%%%%%%%%%%%%%%%%%%%%%%%%

The heavy quark and antiquark have equal mass $M\gg T$ and are treated at leading order in the static expansion. The momentum transfer in the $t$ channel $Q$-$\bar Q$ elastic scattering (see Fig.~\ref{fig:landau-damping}(a)) is
\begin{equation}
	q^\mu=p_1^\mu-p_1'^\mu=p_2'^\mu-p_2^\mu
	=(\omega,\bq),\qquad q=|\bq|,
	\label{eq:transfer}
\end{equation}
where $p_1,p_2$ and $p'_1,p'_2$ label the initial and final heavy-particle momenta. The static limit is $\omega\to0$ at fixed nonzero $q$, so that the exchanged four-momentum is spacelike. The heavy quark state is normalized relativistically 
\begin{equation}
	\avg{\bp',s'|\bp,s}=(2\pi)^3\,2E_{\bp}\,
	\delta^{(3)}(\bp'-\bp)\delta_{s's},
\end{equation}
and the invariant scattering amplitude is defined through
\begin{equation}
	T_{fi}=(2\pi)^4\delta^{(4)}(P_f-P_i)\M_{fi}.
	\label{eq:invariant-amplitude-definition}
\end{equation}
Matching the irreducible exchange amplitude to the Born matrix element of the nonrelativistic interaction gives
\begin{align}
	\overline\M_{fi}
	&=-\sqrt{(2E_1)(2E_2)(2E'_1)(2E'_2)}\,\widetilde V_{fi}
	\nonumber\\
	&=-4M^2\widetilde V_{fi}\,[1+O(\bp^2/M^2)].
	\label{eq:matching}
\end{align}
where the bar refers to the thermal average as defined in Eq.~\eqref{eq:Tthermalaverage} and the tilde denotes a momentum-space kernel. Thus the potential describes the effective interaction of the heavy pair after the medium degrees of freedom have been traced out. At leading static order, the interaction is diagonal and independent of the external heavy-particle spins. After factoring out these spin deltas and projecting onto a fixed color channel, isotropy implies \(\widetilde V(\mathbf q)=\widetilde V(-\mathbf q)\), yielding   
\begin{equation}
	\overline\M_{fi}-\overline\M_{if}^{\,*}
	=-8\ii M^2\,\im\widetilde V_{\mathrm{ex}}(\bq)
	\label{eq:matchingIm}
\end{equation}
for the left hand side of Eq.~\eqref{eq:thermalOT}. The subscript $\mathrm{ex}$ of $\widetilde V_{\mathrm{ex}}(\bq)$ specifies the connected exchange contribution. In the following, we proceed to compute the right hand side of Eq.~\eqref{eq:thermalOT}, which concerns the interference between the amplitudes of the $Q$ and $\bar Q$ scatterings from the bath particles.

%%%%%%%%%%%%%%%%%%%%%%%%%%%%%%%%%%%%%%%%%%%%%%%%%%%%%%%%%%%%%%%
\subsection{Thermal occupation factors from the bath trace}
\label{ssec:thermal}
%%%%%%%%%%%%%%%%%%%%%%%%%%%%%%%%%%%%%%%%%%%%%%%%%%%%%%%%%%%%%%%
To perform the sum over bath states explicitly, we introduce discrete single-particle states for medium particles ({\it i.e.}, thermal light quarks, antiquarks and gluons) that are labeled by $\alpha=(\bk,s,i)$, with momentum $\bk$, spin $s$ and color $i$ for a given flavor and  normalized in a box of volume $\mathcal V$. The relation between box-normalized bath states and relativistically normalized states is
\begin{equation}
	\begin{aligned}
		\ket{\bk,s,i}_{\mathrm{rel}}&=\sqrt{2E_k\mathcal V}\ket{\bk,s,i}_{\mathrm{box}},\\
		\sum_{\bk}&\longrightarrow\mathcal V\intk.
	\end{aligned}
	\label{eq:boxnorm}
\end{equation}

Any medium particles can potentially scatter with $Q$ or $\bar Q$; therefore, the scattering operator $\hat T$ must sum over all possible active medium-particle transitions $\ket\alpha \to \ket\beta$ via
\begin{equation}
	\hat T=\sum_{\alpha,\beta}
	\hat T_{Q,\alpha\beta}\,c_\beta^\dagger c_\alpha,
	\label{eq:Tsc}
\end{equation}
\noindent where $c_\beta^\dagger$ and $c_\alpha$ denote the creation and annihilation operators of the medium particles, respectively, while $\hat T_{Q,\alpha\beta}$ represents the scattering of either heavy constituent as facilitated by medium particles, together with the spectator overlap for the other constituent.

In the following, we proceed by taking the transition mediated by bath quarks as an example. The anti-commutation relation for the creation and annihilation operators is
\begin{equation}
	\{c_\alpha,c_\beta^\dagger\}=\delta_{\alpha\beta},
\end{equation}
with particle number operator $N_\alpha=c_\alpha^\dagger c_\alpha$ for the mode $\alpha$. Then the summation over the bath states on the right hand side of Eq.~\eqref{eq:thermalOT} reads
 
\end{multicols}
\vspace{1pt}
\noindent\begin{minipage}{\textwidth}
\setlength{\abovedisplayskip}{3pt}
\setlength{\belowdisplayskip}{3pt}
\noindent\rule{\dimexpr(\textwidth-\columnsep)/2\relax}{0.4pt}\llap{\rule{0.4pt}{6pt}}\par
\setlength{\medmuskip}{2mu plus 1mu minus 1mu}
\setlength{\thickmuskip}{2.5mu plus 1mu minus 1mu}
\begin{flalign}
&i\sum_a p_a\sum_b T_{nb,fa}^*T_{nb,ia}=i\sum_{\alpha\beta\widetilde\alpha\widetilde\beta}
\bra n\hat T_{Q,\widetilde\alpha\widetilde\beta}\ket f^*
\bra n\hat T_{\bar Q,\alpha\beta}\ket i
\times\sum_a p_a\sum_b
 \bra b c_{\widetilde\beta}^\dagger c_{\widetilde\alpha}\ket a^*
 \bra b c_\beta^\dagger c_\alpha\ket a
=i\sum_{\alpha\beta\widetilde\alpha\widetilde\beta}
 \bra n\hat T_{Q,\widetilde\alpha\widetilde\beta}\ket f^*
&&\nonumber\\
&\quad
 \bra n\hat T_{\bar Q,\alpha\beta}\ket i
\times\delta_{\alpha\widetilde\alpha}
 \delta_{\beta\widetilde\beta}
 \sum_a p_a\bra a c_\alpha^\dagger c_\beta c_\beta^\dagger c_\alpha\ket a
=i\sum_{\alpha\beta}
 \bra n\hat T_{Q,\alpha\beta}\ket f^*
 \bra n\hat T_{\bar Q,\alpha\beta}\ket i
\times\sum_a p_a\bra a
 N_{\alpha}(1-N_{\beta})\ket a.
 &&\label{eq:paTstarT}
\end{flalign}
\noindent\hfill\rlap{\rule[-5.6pt]{0.4pt}{6pt}}\rule{\dimexpr(\textwidth-\columnsep)/2\relax}{0.4pt}\par
\end{minipage}\par
\vspace{3pt}
\begin{multicols}{2}
\noindent where the bath-completeness identity $\sum_b\ket b\bra b=1$ is used in the second step
for a nontrivial transition $\alpha\ne\beta$.

Note that at the order considered, the bath is described by independent fermionic modes:\columnbreak
\begin{equation}
E=\sum_\gamma E_\gamma N_\gamma,\quad
Z=\prod_\gamma Z_\gamma,\quad Z_\gamma=1+e^{-\beta E_\gamma}.
\end{equation}
Then the thermal average in the last step of Eq.~\eqref{eq:paTstarT} is performed,
\end{multicols}
\vspace{1pt}
\noindent\begin{minipage}{\textwidth}
\setlength{\abovedisplayskip}{3pt}
\setlength{\belowdisplayskip}{3pt}
\noindent\rule{\dimexpr(\textwidth-\columnsep)/2\relax}{0.4pt}\llap{\rule{0.4pt}{6pt}}\par
\setlength{\medmuskip}{3mu plus 1mu minus 1mu}
\setlength{\thickmuskip}{3mu plus 1mu minus 1mu}
\begin{flalign}
&\sum_a p_a\bra a N_{\alpha}(1-N_{\beta})\ket a
 =\frac{1}{Z}\sum_{\{N_{\gamma}\}}e^{-\beta\sum_\gamma E_\gamma N_\gamma}N_{\alpha}(1-N_{\beta})
 =\frac{1}{Z_{\alpha}Z_{\beta}}\sum_{N_\alpha,N_\beta=0,1}N_\alpha(1-N_\beta)
 e^{-\beta(E_\alpha N_\alpha+E_\beta N_\beta)} &&\nonumber\\
&\quad=\frac{e^{-\beta E_\alpha}}{(1+e^{-\beta E_\alpha})(1+e^{-\beta E_\beta})}
 =\nf(E_\alpha)[1-\nf(E_\beta)], &&\label{eq:pauli}
\end{flalign}
\noindent\hfill\rlap{\rule[-5.6pt]{0.4pt}{6pt}}\rule{\dimexpr(\textwidth-\columnsep)/2\relax}{0.4pt}\par
\end{minipage}\par
\vspace{3pt}
\begin{multicols}{2} 
\noindent which is indeed the correct thermal factor accounting for the occupation probability of the initial state and the Pauli blocking for the final state for the medium quark transition $\ket\alpha \to \ket\beta$ with $\nf(E)=(e^{\beta E}+1)^{-1}$.  In the second step of the derivation of Eq.~\eqref{eq:pauli}, we have used the fact that all spectator-mode partition functions cancel between the thermal numerator and denominator and thus only the two active modes $\alpha$, $\beta$ survive. 

When dealing with the $Q$ and $\bar Q$ scatterings from bath gluons, similar treatments lead to   
\begin{equation}
\sum_a p_a\bra a N_{\alpha}(1+N_{\beta})\ket a
	=n_B(E_\alpha)[1+n_B(E_\beta)],
	\label{eq:bose}
\end{equation} 
which accounts for the occupation probability of the initial state and the Bose enhancement for the final state for the medium gluon transition $\ket\alpha \to \ket\beta$ with $n_B(E)=(e^{\beta E}-1)^{-1}$. Thus the thermal factors follow from the many-body statistics, rather than being appended to a vacuum cut by hand.

%%%%%%%%%%%%%%%%%%%%%%%%%%%%%%%%%%%%%%%%%%%%%%%%%%%%%%%%%%%%%%%
\subsection{The interference from the light-quark Landau cut}
\label{ssec:landau}
\begin{figure*}[t]
	\centering
	\includegraphics[width=0.92\textwidth]{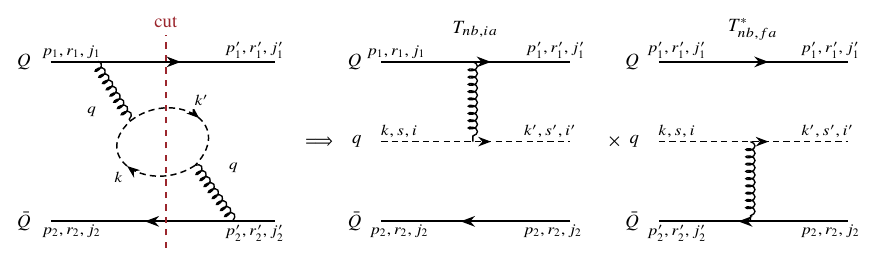}
	\caption{First Landau cut with intermediate heavy-pair state $\ket n=\ket{Q(p_1')\bar{Q}(p_2)}$ and its amplitude product. Solid lines denote the heavy quark and antiquark, dashed lines the light bath quark, and coils the exchanged gluon. The red dashed line is the cut. Each right-hand diagram represents a transition to the same state $n_1b$; the entire second amplitude is complex conjugated. The spectator $\bar Q$ or $Q$ is also indicated. The indices $r_1,r_2$ and $j_1,j_2$ label heavy quark spins and colors, respectively, while $s,i$ label the bath quark spin and color.}
	\label{fig:cut1}
\end{figure*}

\begin{figure*}[t]
	\centering
	\includegraphics[width=0.92\textwidth]{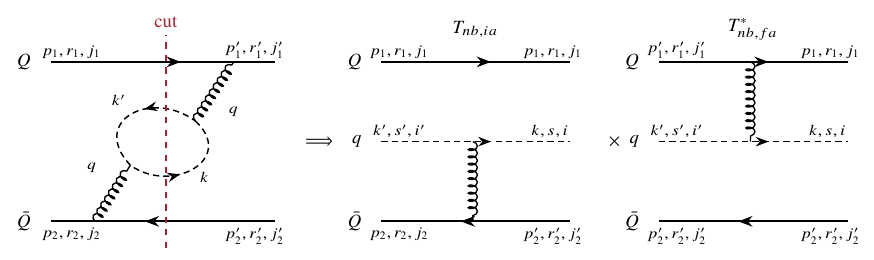}
	\caption{Second Landau cut with intermediate heavy-pair state $\ket n=\ket{Q(p_1)\bar{Q}(p_2')}$and its amplitude product. The bath transition is now $k'\to k$, so that its statistical weight is obtained by interchanging $k$ and $k'$ in that of the first cut.}
	\label{fig:cut2}
\end{figure*}

%%%%%%%%%%%%%%%%%%%%%%%%%%%%%%%%%%%%%%%%%%%%%%%%%%%%%%%%%%%%%%%
For the one-loop light-quark polarization insertion in the gluon exchange amplitude of $Q\bar Q$ scattering from $\ket i=\ket{Q(p_1,r_1,j_1)\bar{Q}(p_2,r_2,j_2)}$ to $\ket f=\ket{Q(p_1',r_1',j_1')\bar{Q}(p_2',r_2',j_2')}$, there are two ways of Landau cuts, as demonstrated in Fig.~\ref{fig:cut1} and Fig.~\ref{fig:cut2}, respectively. 

Upon the first cut with intermediate heavy-pair state $\ket n=\ket{Q(p_1',r_1',j_1')\bar{Q}(p_2,r_2,j_2)}$, one obtains two elastic amplitudes describing the $Q$ and $\bar Q$ scatterings from bath quarks (see right-hand diagrams in Fig.~\ref{fig:cut1} for details of the notations) 
\end{multicols}
\vspace{1pt}
\noindent\begin{minipage}{\textwidth}
\setlength{\abovedisplayskip}{3pt}
\setlength{\belowdisplayskip}{3pt}
% Column-width rules mark where the two-column reading order pauses and resumes.
\noindent\rule{\dimexpr(\textwidth-\columnsep)/2\relax}{0.4pt}\llap{\rule{0.4pt}{6pt}}\par
\begin{flalign}
\bra n\hat T_{Q,\alpha\beta}\ket i&=(2\pi)^4\delta^4(p_1+k-p_1'-k')\M_{Qq}(Q(p_1,r_1,j_1)+q(k,s,i)\to	Q(p_1',r_1',j_1')+q(k',s',i'))\frac{1}{\sqrt{2E_{\bk}\mathcal  V}}\frac{1}{\sqrt{2E_{\bk'}\mathcal V}} &&\label{eq:first-cut-Q-transition}\\[3pt]
\bra n\hat T_{\bar Q,\alpha\beta}\ket f&=(2\pi)^4\delta^4(p_2'+k-p_2-k')\M_{\bar Qq}(\bar Q(p_2',r_2',j_2')+q(k,s,i)\to	\bar Q(p_2,r_2,j_2)+q(k',s',i'))\frac{1}{\sqrt{2E_{\bk}\mathcal V}}\frac{1}{\sqrt{2E_{\bk'}\mathcal V}} &&
\label{eq:first-cut-Qbar-transition}
\end{flalign}
\noindent\hfill\rlap{\rule[-5.6pt]{0.4pt}{6pt}}\rule{\dimexpr(\textwidth-\columnsep)/2\relax}{0.4pt}\par
\end{minipage}\par
\vspace{3pt}
\begin{multicols}{2}
\noindent where we have adopted the single quark states $\ket \alpha=\ket{q(\bk,s,i)}$ and $\ket \beta=\ket{q(\bk',s',i')}$ as well as the conversion between the box-normalization and relativistic normalization (Eq.~\eqref{eq:boxnorm}). The thermal factor (Eq.~\eqref{eq:pauli}) accompanied with this cut is thus 
\begin{equation}
	W_{q,1}(k,k')=\nf(E_{\bk})[1-\nf(E_{\bk'})].
	\label{eq:first-cut-weight}
\end{equation}
Employing the standard Feynman rules, the two invariant amplitudes are
\end{multicols}
\vspace{1pt}
\noindent\begin{minipage}{\textwidth}
\setlength{\abovedisplayskip}{3pt}
\setlength{\belowdisplayskip}{3pt}
\noindent\rule{\dimexpr(\textwidth-\columnsep)/2\relax}{0.4pt}\llap{\rule{0.4pt}{6pt}}\par
\begin{flalign}
  \M_{Qq}
  &= \bar u_Q(p'_1,r_1')\,\ii g\gamma_{\mu}(t_Q^a)_{j'_1j_1}\,u_Q(p_1,r_1)
      \,\delta^{ab}D_R^{\mu\nu}(q)\,
      \bar u_q(k',s')\,\ii g\gamma_{\nu}(t_q^b)_{i'i}\,u_q(k,s) &&\notag\\
  &\simeq(-g^2)2M\delta^{r_1r_1'}(t_Q^a)_{j'_1j_1}\,D_R^{00}(q)\,
      \bar u_q(k',s')\gamma^0u_q(k,s)\,(t_q^a)_{i'i},
      &&\label{eq:first-cut-Qq}\\[5pt]
  \M_{\bar Qq}
  &= -\bar v_Q(p'_2,r_2')\,\ii g\gamma_{\mu}(t_{\bar Q}^a)_{j'_2j_2}\,v_Q(p_2,r_2)
      \,\delta^{ab}D_R^{\mu\nu}(q)\,
      \bar u_q(k',s')\,\ii g\gamma_{\nu}(t_q^b)_{i'i}\,u_q(k,s) &&\notag\\
  &\simeq(+g^2)2M\delta^{r_2r_2'}(t_{\bar Q}^b)_{j'_2j_2}\,D_R^{00}(q)\,
      \bar u_q(k',s')\gamma^0u_q(k,s)\,(t_q^b)_{i'i},
      &&\label{eq:first-cut-Qbarq}
\end{flalign}
\noindent\hfill\rlap{\rule[-5.6pt]{0.4pt}{6pt}}\rule{\dimexpr(\textwidth-\columnsep)/2\relax}{0.4pt}\par
\end{minipage}\par
\vspace{3pt}
\begin{multicols}{2}
\noindent where we have used the static approximation for heavy quark spinors 
\begin{equation}
j_Q^{\mu}=\bar u(p',r')\gamma^{\mu}u(p,r)\simeq2M\delta^{r'r}\delta^{\mu0},
\label{eq:jQmu}
\end{equation}
similarly for $\bar v(p',r')\gamma^{\mu}v(p,r)$. In addition, we have chosen the Coulomb gauge for the retarded gluon propagator 
\begin{equation}
	D_R^{00}(\omega,\bq)=\frac{1}{q^2+\Pi_{00}^R(\omega,q)},~~~
	D_R^{0i}=D_R^{i0}=0.
	\label{eq:coulombprop}
\end{equation}
Combining Eq.~\eqref{eq:jQmu} and Eq.~\eqref{eq:coulombprop}, only the temporal (longitudinal) $00$ component survives in Eq.~\eqref{eq:first-cut-Qq} and Eq.~\eqref{eq:first-cut-Qbarq}. Note that for $\M_{\bar Qq}$ in Eq.~\eqref{eq:first-cut-Qbarq}, an extra minus sign comes from normal ordering of the antiquark operators, $:d_i d_j^\dagger:=-d_j^\dagger d_i$ when constructing the scattering matrix element, so that the overall sign of $\M_{\bar Qq}$ becomes positive. 

Upon summing (averaging) over the intermediate (initial/final) state $Q$ or $\bar Q$ spin indices and summing over the bath quark spin and color indices, the interference between the two elastic amplitudes for the $Q$ and $\bar Q$ scatterings from bath quarks reads
 
\end{multicols}
\vspace{1pt}
\noindent\begin{minipage}{\textwidth}
\setlength{\abovedisplayskip}{3pt}
\setlength{\belowdisplayskip}{3pt}
% Match the full-width layout and reading-order rules of Eq.~\eqref{eq:first-cut-Qbar-transition}.
\noindent\rule{\dimexpr(\textwidth-\columnsep)/2\relax}{0.4pt}\llap{\rule{0.4pt}{6pt}}\par
\begin{flalign}
&\frac{1}{2}\frac{1}{2}\sum_{r_1,r_2,r_1',r_2'}\sum_{s,s',i,i'}\M_{Qq}\M_{\bar Qq}^*
=-g^4\frac{1}{2}\frac{1}{2}\sum_{r_1,r_2,r_1',r_2'}(2M\delta^{r_1r_1'})(2M\delta^{r_2r_2'})|D_R^{00}(q)|^2
  \sum_{a,b}(t_Q^a)_{j'_1j_1}(t_{\bar Q}^b)^*_{j'_2j_2} &&\notag\\
&\quad\times\sum_{i,i'}(t_q^a)_{i'i}(t_q^b)^*_{i'i}
  \sum_{s,s'}\left|\bar u_q(k',s')\gamma^0u_q(k,s)\right|^2
 =-2M^2g^4|D_R^{00}(q)|^2\,L^{00}(k,k')
  \sum_a(t^a_Q)_{j'_1j_1}(t^a_{\bar Q})^*_{j'_2j_2}.
  &&\label{eq:first-cut-interference}
\end{flalign}
\noindent\hfill\rlap{\rule[-5.6pt]{0.4pt}{6pt}}\rule{\dimexpr(\textwidth-\columnsep)/2\relax}{0.4pt}\par
\end{minipage}\par
\vspace{3pt}
\begin{multicols}{2}
\noindent where the bath quark color sum
\begin{align}
	\sum_{i,i'}(t^a)_{i'i}(t^b)^*_{i'i}=T_F\delta^{ab}, \qquad T_F=\frac12,
	\label{eq:qcolortrace}
\end{align}
while the external heavy colors remain uncontracted, and the bath quark spin sum
\begin{align}
	L^{00}(k,k')
	&\equiv\sum_{s,s'}\left|\bar u_q(k',s')\gamma^0u_q(k,s)\right|^2
	\nonumber\\*
	&=\tr(\slashed{k}'\gamma^0\slashed{k}\gamma^0)
	=4(kk'+\bk\cdot\bk').
	\label{eq:spintrace} 
\end{align}

We now project the $Q$ and $\bar Q$ participating in the elastic scattering onto a color singlet, with color wave functions $\Psi^{(s)}_{j_1j_2}=\delta^{j_1j_2}/\sqrt{N_c}$ in the initial state and $\delta^{j_1'j_2'}/\sqrt{N_c}$ in the final state, respectively (see the left-hand diagram in Fig.~\ref{fig:cut1}). The remaining $Q$ and $\bar Q$ colors indices in Eq.~\eqref{eq:first-cut-interference} is then contracted with these wave functions
\begin{equation}
	\begin{aligned}
		\mathcal C_{s}
		&\equiv\sum_{a}\sum_{j_1,j_2,j'_1,j'_2}
		(t^a_Q)_{j'_1j_1}(t^a_{\bar Q})^*_{j'_2j_2}
		\frac{\delta_{j_1j_2}}{\sqrt{N_c}}
		\frac{\delta_{j'_1j'_2}}{\sqrt{N_c}}\\
		&=\frac{1}{N_c}\sum_a\tr(t^at^a)=\frac{N_c^2-1}{2N_c}=C_F. 
	\end{aligned}
	\label{eq:singletindices}
\end{equation}

Now combining Eqs.~\eqref{eq:first-cut-interference}, ~\eqref{eq:spintrace} and ~\eqref{eq:singletindices} gives the interference between the amplitudes of $Q$ and $\bar Q$ (in the color singlet) scatterings from bath quarks from the first cut
\begin{equation}
 \sum\M_{Qq}\M_{\bar Qq}^*=-2C_Fg^4M^2|D_R^{00}|^2\,4(kk'+\bk\cdot\bk').
 \label{eq:interference-first-cut}
\end{equation}

For the second cut with intermediate heavy-pair state $\ket n=\ket{Q(p_1,r_1,j_1)\bar{Q}(p_2',r_2',j_2')}$, the $\bar Q$ scattering from a bath quark occurs first, which is followed by the $Q$ scattering from the same bath quark; these two elastic amplitudes are (see right-hand diagrams in Fig.~\ref{fig:cut2} for details of the notations)
\end{multicols}
\vspace{1pt}
\noindent\begin{minipage}{\textwidth}
\setlength{\abovedisplayskip}{3pt}
\setlength{\belowdisplayskip}{3pt}
\noindent\rule{\dimexpr(\textwidth-\columnsep)/2\relax}{0.4pt}\llap{\rule{0.4pt}{6pt}}\par
\begin{flalign}
\bra n\hat T_{\bar Q,\alpha\beta}\ket i
&=(2\pi)^4\delta^4(p_2+k'-p_2'-k)\M_{\bar Qq}\bigl(\bar Q(p_2,r_2,j_2)+q(k',s',i')\to\bar Q(p'_2,r'_2,j'_2)+q(k,s,i)\bigr)
\frac{1}{\sqrt{2E_{\bk}\mathcal V}}\frac{1}{\sqrt{2E_{\bk'}\mathcal V}} &&\label{eq:second-cut-Qbar-transition}\\[3pt]
\bra n\hat T_{Q,\alpha\beta}\ket f
&=(2\pi)^4\delta^4(p_1'+k'-p_1-k)\M_{Qq}\bigl(Q(p'_1,r'_1,j'_1)+q(k',s',i')\to Q(p_1,r_1,j_1)+q(k,s,i)\bigr)
\frac{1}{\sqrt{2E_{\bk}\mathcal V}}\frac{1}{\sqrt{2E_{\bk'}\mathcal V}} &&
\label{eq:second-cut-Q-transition}
\end{flalign}
\noindent\hfill\rlap{\rule[-5.6pt]{0.4pt}{6pt}}\rule{\dimexpr(\textwidth-\columnsep)/2\relax}{0.4pt}\par
\end{minipage}\par
\vspace{3pt}
\begin{multicols}{2}
\noindent Here we've labeled the single-quark states $\ket\alpha=\ket{q(\bk',s',i')}$ and $\ket\beta=\ket{q(\bk,s,i)}$ for the bath transition $\ket\alpha \to\ket\beta$, which is thus accompanied by the thermal factor according to Eq.~\eqref{eq:pauli}  
\begin{equation}
  W_{q,2}(k,k')=\nf(E_{\bk'})[1-\nf(E_{\bk})].
  \label{eq:second-cut-weight}
\end{equation}

One notes that the invariant amplitude $\M_{Qq}$ in the first cut (Eq.~\eqref{eq:first-cut-Q-transition}) and that in the second cut (Eq.~\eqref{eq:second-cut-Q-transition}) differ only by swapping their initial and final states. The same is true for the other invariant amplitude $\M_{\bar Q q}$ (Eq.~\eqref{eq:first-cut-Qbar-transition} and Eq.~\eqref{eq:second-cut-Qbar-transition}. By making use of the parity and time-reversal symmetry which swap the momenta and spins (with flipped sign) for the initial and final state, the spin summed (averaged) interference between these two amplitudes $\sum\M_{\bar Q q}\M_{Qq}^*$ for the second cut thus remains the same as for the first cut $\sum\M_{Qq}\M_{\bar Q q}^*$ (Eq.~\eqref{eq:interference-first-cut}).

%%%%%%%%%%%%%%%%%%%%%%%%%%%%%%%%%%%%%%%%%%%%%%%%%%%%%%%%%%%%%%%
\subsection{The imaginary potential in the static HTL limit}
\label{ssec:staticHTL}
%%%%%%%%%%%%%%%%%%%%%%%%%%%%%%%%%%%%%%%%%%%%%%%%%%%%%%%%%%%%%%%

Now combining the results from two ways of Landau cuts including their respective thermal factors (Eqs.~\eqref{eq:first-cut-weight} and~\eqref{eq:second-cut-weight}), plugging all pertinent ingredients Eqs.~\eqref{eq:first-cut-Q-transition}, \eqref{eq:first-cut-Qbar-transition}, \eqref{eq:interference-first-cut}, ~\eqref{eq:second-cut-Qbar-transition}, \eqref{eq:second-cut-Q-transition}, into Eq.~\eqref{eq:paTstarT}, and matching the result to the absorptive part of the non-forward amplitude in Eq.~\eqref{eq:matchingIm}, we arrive at the exchange imaginary $Q\bar Q$ potential
\end{multicols}
\vspace{1pt}
\noindent\begin{minipage}{\textwidth}
\setlength{\abovedisplayskip}{3pt}
\setlength{\belowdisplayskip}{3pt}
\noindent\rule{\dimexpr(\textwidth-\columnsep)/2\relax}{0.4pt}\llap{\rule{0.4pt}{6pt}}\par
\begin{flalign}
\im\widetilde V_{\mathrm{ex}}(\bq)
&=2N_fC_Fg^4\sum_{\bk}\sum_{\bk'}
\frac{1}{2E_{\bk}\mathcal V}\frac{1}{2E_{\bk'}\mathcal V}
(2\pi)^4\delta^{(4)}(p_1+k-p'_1-k')
\left|D_R^{00}(q)\right|^2(kk'+\bk\cdot\bk') &&\notag\\[3pt]
&\quad\times\Bigl[W_{q,1}(k,k')+W_{q,2}(k,k')\Bigr],
&&\label{eq:exchange-imaginary-box}
\end{flalign}
\noindent\hfill\rlap{\rule[-5.6pt]{0.4pt}{6pt}}\rule{\dimexpr(\textwidth-\columnsep)/2\relax}{0.4pt}\par
\end{minipage}\par
\vspace{3pt}
\begin{multicols}{2}
\noindent where the heavy quark mass factors cancel out and a degeneracy factor $2N_f$ accounting for flavors and bath antiquarks is inserted. In deriving Eq.~\eqref{eq:exchange-imaginary-box}, we have taken out a common energy-momentum conserving $\delta^4(p_1+p_2-p_1'-p_2')$ function on both sides.

Converting the box-normalization summation over discrete momentum into an integration using Eq.~\eqref{eq:boxnorm} and integrating out the 3-dimension $\delta$ function, one obtains
\end{multicols}
\vspace{1pt}
\noindent\begin{minipage}{\textwidth}
\setlength{\abovedisplayskip}{3pt}
\setlength{\belowdisplayskip}{3pt}
\noindent\rule{\dimexpr(\textwidth-\columnsep)/2\relax}{0.4pt}\llap{\rule{0.4pt}{6pt}}\par
\begin{flalign}
\im\widetilde V_{\mathrm{ex}}(\bq)
&=\frac{N_fC_Fg^4}{8\pi^2}\int\!\dd^3\bk\,
\delta\!\left(E_{\bp_1}+E_{\bk}-E_{\bp'_1}-E_{\bk'}\right)
\frac{kk'+\bk\cdot\bk'}{E_{\bk}E_{\bk'}}
\left|D_R^{00}(q)\right|^2\Bigl[W_{q,1}(k,k')+W_{q,2}(k,k')\Bigr] .
&&\label{eq:exchange-imaginary-integrated}
\end{flalign}
\noindent\hfill\rlap{\rule[-5.6pt]{0.4pt}{6pt}}\rule{\dimexpr(\textwidth-\columnsep)/2\relax}{0.4pt}\par
\end{minipage}\par
\vspace{3pt}
\begin{multicols}{2}

 Now we are in a position to implement the leading order static limit, in which $q^0=\omega=0$ and the longitudinal hard-thermal-loop (HTL) gluon propagator takes the form
 \begin{equation}
 D^{00}_R(\omega=0,\bq)=\frac{1}{q^2+m_D^2},
 \label{eq:D00HTL}
 \end{equation}
 with the Debye screening mass $m_D^2=\Pi_{00}^R(0,q)=g^2T^2(N_c/3+N_f/6)$ for massless bath particles. Also in this limit, there's no recoil on $Q$ or $\bar Q$ such that $E_{\bp_1'}\simeq E_{\bp_1}$. In the HTL approximation, the bath quark momentum $k\sim T$; for soft momentum transfer $q\ll T$, $\bk'=\bk+\bq$, $k'=k+\hat{\bk}\cdot\bq+O(q^2/k)$, giving
 \begin{align}
 &\delta(E_{\bp_1}+E_{\bk}-E_{\bp_1'}-E_{\bk'})\simeq\delta(k'-k)\simeq\delta(\widehat{\bk}\cdot\bq),\nonumber\\
 &\frac{kk'+\bk\cdot\bk'}{kk'}\simeq2,\nonumber\\
& W_{q,1}(k,k')+W_{q,2}(k,k')\simeq2\nf(k)[1-\nf(k)].
 \label{eq:HTL}		
 \end{align}
 
Substituting Eqs.~\eqref{eq:D00HTL} and ~\eqref{eq:HTL} into Eq.~\eqref{eq:exchange-imaginary-integrated} and using
\begin{align}
\int_{-1}^{1}\dd\cos\theta\,\delta(q\cos\theta)=\frac1q,\nonumber \\
 \int_0^\infty\dd k\,k^2\nf(k)[1-\nf(k)]=\frac{\pi^2T^3}{6},
 \end{align}
the exchange imaginary $Q\bar Q$ potential from the light-quark Landau cut finally reads
\begin{equation}
	\im\widetilde V_{\mathrm{ex},s}^{(q)}(\bq)
	=\frac{g^2C_F\pi Tm_{D,q}^2}{q(q^2+m_D^2)^2}.
	\label{eq:ImVq}
\end{equation}
with $m_{D,q}^2=g^2T^2N_f/6$ being the light-quark contribution to the Debye screening mass squred.

%%%%%%%%%%%%%%%%%%%%%%%%%%%%%%%%%%%%%%%%%%%%%%%%%%%%%%%%%%%%%%%%

\subsection{Thermal-gluon contribution}
\label{sec:gluons}

\begin{figure*}[t]
\centering
\includegraphics[width=0.92\textwidth]{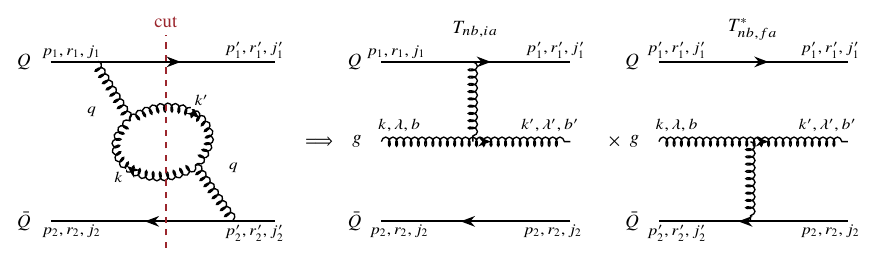}
\caption{First Landau cut of the one-loop gluon self-energy insertion and its amplitude interference product, with the heavy-state assignments of Fig.~\ref{fig:cut1}. The bath gluon $g$ carries incoming momentum $k$, polarization $\lambda$ and color $b$ and outgoing labels $k',\lambda',b'$.}
\label{fig:gluon-cut1}
\end{figure*}

\begin{figure*}[t]
\centering
\includegraphics[width=0.92\textwidth]{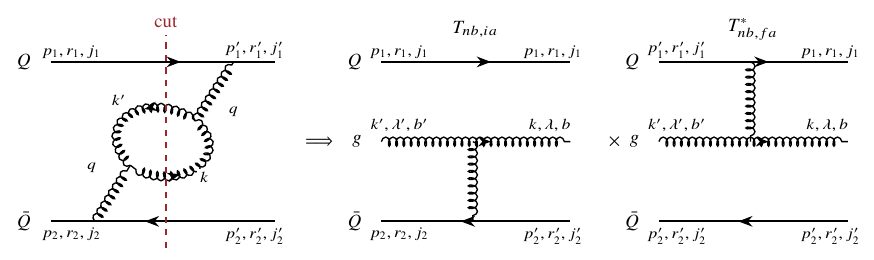}
\caption{Second Landau cut of the one-loop gluon self-energy insertion and its amplitude interference product, with the heavy-state assignments of Fig.~\ref{fig:cut2}. The bath-gluon  now makes the transition from $k',\lambda',b'$ to $k,\lambda,b$.}
\label{fig:gluon-cut2}
\end{figure*}

For the one-loop gluon self-energy insertion in the spacelike gluon exchange amplitude of $Q\bar Q$ scattering, the two gluonic Landau-cut assignments and their amplitude interference products are shown in Figs.~\ref{fig:gluon-cut1} and \ref{fig:gluon-cut2}, in parallel with Figs.~\ref{fig:cut1} and \ref{fig:cut2}.


Involving a three-gluon vertex, the invariant amplitude for $Q$ scattering from a bath gluon reads (see notations in the right-hand diagrams in Fig.~\ref{fig:gluon-cut1})
\end{multicols}
\vspace{1pt}
\noindent\begin{minipage}{\textwidth}
\setlength{\abovedisplayskip}{3pt}
\setlength{\belowdisplayskip}{3pt}
\noindent\rule{\dimexpr(\textwidth-\columnsep)/2\relax}{0.4pt}\llap{\rule{0.4pt}{6pt}}\par
\begin{flalign}
	\M_{Qg}
	&=\bar u_Q(p'_1,r_1')\,\ii g\gamma_{\mu}(t_Q^a)_{j'_1j_1}\,u_Q(p_1,r_1)
	D_R^{\mu\nu}(q)\,(-i)gf^{abb'}\epsilon^{\alpha}(k,\lambda)\epsilon^{*\beta}(k',\lambda') &&\notag\\[3pt]
	&\quad\times\bigl[g_{\nu\alpha}(q-k)_{\beta}+g_{\alpha\beta}(k+k')_{\nu}+g_{\beta\nu}(-k'-q)_{\alpha}\bigr].
	&&\label{eq:first-cut-Qg}
\end{flalign}
\noindent\hfill\rlap{\rule[-5.6pt]{0.4pt}{6pt}}\rule{\dimexpr(\textwidth-\columnsep)/2\relax}{0.4pt}\par
\end{minipage}\par
\vspace{3pt}
\begin{multicols}{2}
\noindent Using the same static approximation (Eq.~\eqref{eq:jQmu}) and working in the Coulomb gauge (Eq.~\eqref{eq:coulombprop}) as in the case of the light-quark Landau cuts alongside physical transverse gluon polarizations
\begin{equation}
	\epsilon^{\alpha}(k,\lambda)=(0,\bm\epsilon_{\lambda}(\bm k)),
\end{equation}
one ends up with the heavy source coupling to the screened temporal gluon field
\begin{equation}
\begin{aligned}
	\M_{Qg}&\simeq(-g^2)2M\delta^{r_1r_1'}D_R^{00}(t_Q^a)_{j'_1j_1}f^{abb'}\\
	&\quad\times(k^0+k'^0)\bm\epsilon_{\lambda'}^*(\bk')\cdot\bm\epsilon_{\lambda}(\bk).
\end{aligned}
	\label{eq:gluonamplitudes}
\end{equation}
Similarly the invariant amplitude for $\bar Q$ scattering from a bath gluon reads 
\begin{equation} 
\begin{aligned}
	\M_{\bar Q g}&\simeq(+g^2)2M\delta^{r_2r_2'}D_R^{00}(t_{\bar Q}^c)_{j'_2j_2}f^{cbb'}\\
	&\quad\times(k^0+k'^0)\bm\epsilon_{\lambda'}^*(\bk')\cdot\bm\epsilon_{\lambda}(\bk)
\end{aligned}
\end{equation}
where the relative sign is again from the antiquark normal-ordering.

Using the bath gluon color and polarization sums   
\begin{align}
 &\sum_{b,b'}f^{abb'}f^{cbb'}=C_A\delta^{ac},\\
 &L_g^{00}\equiv\sum_{\lambda,\lambda'}\bigl|(k^0+k'^0)\bm\epsilon^*_{\lambda'}(\bk')\cdot\bm\epsilon_{\lambda}(\bk)\bigr|^2\nonumber\\
 &\phantom{L_g^{00}}=(k^0+k'^0)^2[1+(\widehat{\bk}\cdot\widehat{\bk}')^2].
 \label{eq:gluonpolsum}
\end{align}
and contracting with the singlet wave functions as done in Eq.~\eqref{eq:singletindices} finally gives the heavy-pair spin-averaged interference amplitude product for $Q$ and $\bar Q$ scatterings from bath gluons for the first cut (Fig.~\ref{fig:gluon-cut1})
\begin{align}
	\sum\M_{Qg}\M_{\bar Qg}^*
	=&-4M^2g^4|D_R^{00}|^2 C_AC_F \nonumber \\
	&\times(k^0+k'^0)^2[1+(\widehat{\bk}\cdot\widehat{\bk}')^2].
	\label{eq:gluoninterference}
\end{align}
which should be supplemented with the thermal factor accounting for bath gluons' occupation and Bose enhancement $W_{g,1}=n_B(E(\bk))[1+n_B(E(\bk')]$ with $n_B(E(\bk))=(e^{E(\bk)/T}-1)^{-1}$.
	
In a similar manner, one can obtain the interference amplitude product $\sum\M_{\bar Qg}\M_{Qg}^*$ for the second cut (Fig.~\ref{fig:gluon-cut2}) supplemented with the thermal factor $W_{g,2}=n_B(E(\bk'))[1+n_B(E(\bk)]$. Combining results from these two cuts and going through the static and HTL approximations leading to Eq.~\eqref{eq:ImVq}, we finally arrive at the exchange imaginary $Q\bar Q$ potential from the thermal-gluon Landau cut

\begin{equation}
	\im\widetilde V_{\mathrm{ex},s}^{(g)}(\bq)
	=\frac{g^2C_F\pi Tm_{D,g}^2}{q(q^2+m_D^2)^2}.
	\label{eq:ImVg}
\end{equation}
with $m_{D,g}^2=g^2T^2C_A/3$ ($C_A=N_c$ for $SU(N_c)$) being the gluon contribution to the Debye screening mass squared.

\section{The full imaginary potential with single-quark self-energies}
\label{sec:full-ImV}
%%%%%%%%%%%%%%%%%%%%%%%%%%%%%%%%%%%%%%%%%%%%%%%%%%%%%%%%%%%%%%%%

%%%%%%%%%%%%%%%%%%%%%%%%%%%%%%%%%%%%%%%%%%%%%%%%%%%%%%%%%%%%%%%
\subsection{Single-heavy-quark self-energy}
\label{ssec:selfenergy}
%%%%%%%%%%%%%%%%%%%%%%%%%%%%%%%%%%%%%%%%%%%%%%%%%%%%%%%%%%%%%%%
While the interference amplitude products of $Q$ and $\bar Q$ scatterings from bath particles yield the exchange $Q\bar Q$ imaginary potential as discussed in Sec.~\ref{ssec:landau}, Sec.~\ref{ssec:staticHTL} and Sec.~\ref{sec:gluons}, the individual scattering amplitude squared gives rise to single heavy-quark self-energies~\cite{Beraudo2008}.

At leading order in the static expansion, the heavy-light quark scattering amplitude squared, with initial (final) state spin-color degeneracies averaged (summed over) from Eq.~\eqref{eq:first-cut-Qq}, reads 
\begin{equation}
	\overline{|\M_{Qq}|^2}
	=\frac{4M^2g^4C_F}{N_c}|D_R^{00}(0,q)|^2
	(kk'+\bk\cdot\bk').
	\label{eq:rateamplitude}
\end{equation}
Denoting $\dd\Pi_k=\dd^3\bk/(2\pi)^3 2E_k$ and including the incoming light-quark's spin, color, anti-particle degeneracies, the inclusive scattering rate for a given heavy quark is
\end{multicols}
\vspace{1pt}
\noindent\begin{minipage}{\textwidth}
\setlength{\abovedisplayskip}{3pt}
\setlength{\belowdisplayskip}{3pt}
\noindent\rule{\dimexpr(\textwidth-\columnsep)/2\relax}{0.4pt}\llap{\rule{0.4pt}{6pt}}\par
\begin{flalign}
\Gamma_Q^{(q)}&=\frac{4N_cN_f}{2M}\int\dd\Pi_k\,\dd\Pi_{k'}\,\dd\Pi_{p'} (2\pi)^4\delta^{(4)}(p+k-p'-k')
\nf(k)[1-\nf(k')]\,\overline{|\M_{Qq}|^2} &&\notag\\
&\simeq N_fC_Fg^4\intq|D_R^{00}(0,q)|^2\intk2\pi\delta(k-k')\frac{kk'+\bk\cdot\bk'}{kk'}\nf(k)[1-\nf(k')]. &&\label{eq:rate}
\end{flalign}
\noindent\hfill\rlap{\rule[-5.6pt]{0.4pt}{6pt}}\rule{\dimexpr(\textwidth-\columnsep)/2\relax}{0.4pt}\par
\end{minipage}\par
\vspace{3pt}
\begin{multicols}{2}
\noindent where we have used $2E_{p'}\simeq2M$ and changed variables from $(\bk,\bk')$ to $(\bk,\bq)$, with $\bq=\bk'-\bk$ after fixing $\bp'=\bp+\bk-\bk'$ from the spatial $\delta$ function. The same static HTL reduction as in Eq.~\eqref{eq:HTL} therefore yields
\begin{equation}
	\Gamma_Q^{(q)}=\intq\frac{g^2C_F\pi Tm_{D,q}^2}{q(q^2+m_D^2)^2}  =\frac{g^2C_FT}{4\pi}\frac{\mq^2}{m_D^2}.
	\label{eq:singlewidth}
\end{equation}
 
Applying the same static HTL reduction to the squared heavy-quark–-gluon scattering amplitude gives the gluonic contribution to the scattering rate for a single heavy source 
\begin{equation}
 	\Gamma_Q^{(g)}=\intq\frac{g^2C_F\pi Tm_{D,g}^2}{q(q^2+m_D^2)^2} 
 =\frac{g^2C_FT}{4\pi}\frac{m_{D,g}^2}{m_D^2},
 \label{eq:gluonwidth}
\end{equation}
Combining the light-quark and gluon contributions, the imaginary part of the self-energy for either $Q$ or $\bar Q$ is
\begin{align}
 &\im E_Q=\im E_{\bar Q}=(-\frac{\Gamma_Q^{(q)}}{2})+ (- \frac{\Gamma_Q^{(g)}}{2})\nonumber\\
 &=-\frac{1}{2}\intq\frac{g^2C_F\pi Tm_{D}^2}{q(q^2+m_D^2)^2}=-\frac{g^2C_FT}{8\pi} 
 \label{eq:imagself}
\end{align}
where $m_D^2=m_{D,q}^2+m_{D,g}^2$ is used.

%%%%%%%%%%%%%%%%%%%%%%%%%%%%%%%%%%%%%%%%%%%%%%%%%%%%%%%%%%%%%%%
\subsection{Full imaginary potential in the color-singlet channel}
\label{ssec:ImV-r-singlet}
%%%%%%%%%%%%%%%%%%%%%%%%%%%%%%%%%%%%%%%%%%%%%%%%%%%%%%%%%%%%%%%

The full imaginary potential of the $Q\bar Q$ system is defined as the sum of the exchange imaginary potential arising from the interference between scattering amplitudes of $Q$ and $\bar Q$ from bath quarks/gluons and the imaginary self-energies of single $Q$ and $\bar Q$, and thus measures the decay of the whole system. 

We therefore first combine the interference imaginary potentials from light-quark and gluon cuts, Eq.~\eqref{eq:ImVq} and Eq.~\eqref{eq:ImVg}
\begin{align}
 \im\widetilde V_{\mathrm{ex},s}(q)&\equiv\im\widetilde V_{\mathrm{ex},s}^{(q)}(q)
                 +\im\widetilde V_{\mathrm{ex},s}^{(g)}(q)\nonumber \\
 &=\frac{g^2C_F\pi Tm_D^2}{q(q^2+m_D^2)^2}.
\label{eq:fullImV}
\end{align}
Its Fourier transform supplemented with the $Q$ and $\bar Q$ imaginary self-energies finally yields the full color-singlet $Q\bar Q$ imaginary potential in coordinate space 
\begin{align}
\im V_{s}(r)&=\intq \im \widetilde V_{\mathrm{ex},s}(q) e^{\ii\bq\cdot\br} +\im E_Q+\im E_{\bar Q}\nonumber\\
&=-g^2C_F\intq(1- e^{\ii\bq\cdot\br})\frac{\pi Tm_D^2}{q(q^2+m_D^2)^2}\nonumber\\
&\equiv-\frac{g^2C_FT}{4\pi}\phi(m_Dr).
	\label{eq:fullImV-cooridinate-space}
\end{align}
Taking the angular average of $e^{\ii\bq\cdot\br}$ gives $\sin(qr)/(qr)$; with $x=m_Dr$ and $z=q/m_D$, we then have
\begin{equation}
	\phi(x)=2\int_0^\infty\!\dd z\,\frac{z}{(1+z^2)^2}
	\left[1-\frac{\sin(zx)}{zx}\right].
	\label{eq:phi}
\end{equation}
Eq.~\eqref{eq:fullImV-cooridinate-space} is exactly the imaginary part of the ``static potential'' derived from different perspectives in Refs.~\cite{Laine2007,Beraudo2008} for a heavy quark-antiquark pair propagating in the QGP.
 
One notes that at large separation, $\phi(\infty)=1$ such that $\im V_s(\infty)=-g^2C_FT/4\pi=\im E_Q+\im E_{\bar Q}$, implying that the imaginary $Q\bar Q$ potential reduces to the independent damping of two widely separated heavy constituents. In contrast, at small separation $r\to 0$, $\phi(0)=0$ and the imaginary potential in the singlet channel vanishes: a compact $Q\bar Q$ singlet is protected by destructive interference between the scattering from $Q$ and the scattering from $\bar Q$ as assisted by bath particles. The latter can be better understood from constructing the interference by placing $Q$ and $\bar Q$ at $\br/2$ and $-\br/2$, respectively,
\begin{align}
	&~~|\M_{Qq}e^{i\bq\cdot\br/2}+\M_{\bar Qq}e^{-i\bq\cdot\br/2}|^2\nonumber\\
	&=|\M_{Qq}|^2 + |\M_{\bar Qq}|^2 + 2\M_{Qq}\M_{\bar Qq}^*\cos(\bq\cdot\br).	\label{eq:constructing-interference}
\end{align}
But $|\M_{Qq}|^2=|\M_{\bar Qq}|^2$ (Eq.~\eqref{eq:rateamplitude}) after undoing the average over the incoming light-quark's spin-color degeneracy ({\it i.e.} times $2\cdot3=6$) becomes equal to $\M_{Qq}\M_{\bar Qq}^*$ (Eq.~\eqref{eq:interference-first-cut}, which is real) in magnitude but with opposite sign, which means, at $r=0$, the maximum interference contribution exactly cancels the single heavy source self-energies, thereby leading to a vanishing decay width $\Gamma_{Q\bar Q,s}(r)=-2\im V_s(r)$ for the singlet $Q\bar Q$ system.

\subsection{Full imaginary potential in the color-octet channel}
\label{ssec:ImV-r-octet}
%%%%%%%%%%%%%%%%%%%%%%%%%%%%%%%%%%%%%%%%%%%%%%%%%%%%%%%%%%%%%%%
For a fixed adjoint color label $c$, the normalized octet wave function is
\begin{equation}
\Psi^{(o,c)}_{ij}=\sqrt{2}(t^c)_{ij},\qquad\sum_{ij}|\Psi^{(o,c)}_{ij}|^2=1.
\label{eq:octetwave}
\end{equation}
%which is orthogonal to the color singlet wave function
%\begin{equation}
%\sum_{ij}\Psi^{(o,c)*}_{ij}\Psi^{(s)}_{ij}=0.
%\end{equation}
Using these octet wave functions in place of the singlet one in Eq.~\eqref{eq:singletindices} to contract the $Q$ and $\bar Q$ colors in both light-quark and gluonic cuts with the help of the identity $\sum_atr(t^at^bt^at^c)=-\delta^{bc}/4N_c$ and summing over both sectors therefore gives the interference imaginary potential in the color octet channel 
\begin{equation}
\begin{aligned}
 \im\widetilde V_{\mathrm{ex},o}(q)
 &=\im\widetilde V_{\mathrm{ex},o}^{(q)}(q)
   +\im\widetilde V_{\mathrm{ex},o}^{(g)}(q)\\
 &=-\frac{g^2}{2N_c}\frac{\pi Tm_D^2}{q(q^2+m_D^2)^2}.
\end{aligned}
\label{eq:octetexchange}
\end{equation}
Its Fourier transform supplemented with the heavy source imaginary self-energies (Eq.~\eqref{eq:imagself}) gives the full color-octet $Q\bar Q$ imaginary potential in coordinate space 
\begin{equation}
 \im V_o(r)=\frac{g^2T}{4\pi}\left[\frac{\phi(m_Dr)}{2N_c}-\frac{N_c}{2}\right],
\label{eq:V-r-octet}
\end{equation}
which is negative definite.

While the large separation limit of the octet imaginary potential $\im V_o(\infty)=-g^2C_FT/4\pi$ is still given by the sum of independent damping rates of two heavy sources, the small separation limit $\im V_o(0)=-g^2N_cT/8\pi$ is non-vanishing, simply because the octet $Q\bar Q$ carries a net adjoint charge and thus has a finite damping rate even at zero separation via color interactions with medium particles.

%%%%%%%%%%%%%%%%%%%%%%%%%%%%%%%%%%%%%%%%%%%%%%%%%%%%%%%%%%%%%%%%
\section{Summary}
\label{sec_sum}
%%%%%%%%%%%%%%%%%%%%%%%%%%%%%%%%%%%%%%%%%%%%%%%%%%%%%%%%%%%%%%%%
In this work, we have developed an alternative derivation of the imaginary heavy quark-antiquark potential in a weakly coupled QGP. The approach is built upon the in-medium version of the generalized optical theorem, which provides a direct route from elastic scattering of $Q$ and $\bar Q$ with bath quarks and gluons to the Landau-cut contribution to the imaginary potential. Three ingredients are distinct in this construction. The thermal trace fixes Pauli blocking or Bose enhancement. The two cut assignments produce the interference of scatterings of $Q$ and $\bar Q$ from bath particles that determines how the pair damping depends on the separation. The diagonal scattering probabilities determine the imaginary self-energies that complete the interference contribution. For the color singlet, the interference kernel is offset by the negative single-source self-energies, yielding an imaginary potential that vanishes at zero separation. For the octet, the exchange interference changes sign while the single-source terms remain unchanged, always giving a non-vanishing result. Our work thus offers a transparent window into the origin of the imaginary $Q\bar Q$ potential: it arises essentially from a thermal-bath-assisted interference mechanism, supplemented by the independent damping of single heavy sources.”

Finite heavy-quark recoil, spin-dependent interactions, nonzero chemical potentials, finite binding-energy effect~\cite{Brambilla2008,Blaizot:2021xqa} and next-to-leading order corrections (as recently investigated in~\cite{Carrington:2024ize}) remain outside the approximation adopted here and may be further pursued within the present framework.

\section*{Acknowledgments}
This work was supported by the National Natural Science Foundation of China (NSFC) under Grant No. 12475141 and the Fundamental Research Funds for the Central Universities under Grant No. 30925020109.

\begingroup
\raggedright

\endgroup
\end{multicols}

\small
\begin{thebibliography}{99}
	
\bibitem{Eichten1978}
E.~Eichten, K.~Gottfried, T.~Kinoshita, K.~D.~Lane and T.-M.~Yan,
%\emph{Charmonium: The model},
Phys. Rev. D \textbf{17} (1978) 3090.
%\href{https://doi.org/10.1103/PhysRevD.17.3090}{doi:10.1103/PhysRevD.17.3090}.
	
\bibitem{Matsui1986}
 T.~Matsui and H.~Satz,
% \emph{$J/\psi$ suppression by quark-gluon plasma formation},
 Phys. Lett. B \textbf{178} (1986) 416--422.
% \href{https://doi.org/10.1016/0370-2693(86)91404-8}{doi:10.1016/0370-2693(86)91404-8}.

 
 
 \bibitem{Brambilla:2016wgg}
 N.~Brambilla, M.~A.~Escobedo, J.~Soto and A.~Vairo,
% \emph{Quarkonium suppression in heavy-ion collisions: an open quantum system approach},
 Phys. Rev. D \textbf{96}, no.3 (2017) 034021. 
%\href{https://doi:10.1103/PhysRevD.96.034021}{doi:10.1103/PhysRevD.96.034021}.


\bibitem{Blaizot:2017ypk}  
J.~P.~Blaizot and M.~A.~Escobedo,
% \emph{Quantum and classical dynamics of heavy quarks in a quark-gluon plasma},
JHEP \textbf{06} (2018) 034 .
%\href{https://doi:10.1007/JHEP06(2018)034}{doi:10.1007/JHEP06(2018)034}.
 
 
 \bibitem{Yao:2021lus}
 X.~Yao,
% \emph{Open quantum systems for quarkonia},
 Int. J. Mod. Phys. A \textbf{36}, no.20 (2021) 2130010.
%\href{https://doi:10.1142/S0217751X21300106}{doi:10.1142/S0217751X21300106}.
  
 

\bibitem{Akamatsu2022}
 Y.~Akamatsu,
% \emph{Quarkonium in quark--gluon plasma: Open quantum system approaches re-examined},
 Prog. Part. Nucl. Phys. \textbf{123} (2022) 103932.
% \href{https://doi.org/10.1016/j.ppnp.2021.103932}{doi:10.1016/j.ppnp.2021.103932}.
 
 
 
\bibitem{Laine2007}
	M.~Laine, O.~Philipsen, P.~Romatschke and M.~Tassler,
%	\emph{Real-time static potential in hot QCD},
	JHEP \textbf{03} (2007) 054.
%	\href{https://doi.org/10.1088/1126-6708/2007/03/054}{doi:10.1088/1126-6708/2007/03/054},
%	\href{https://arxiv.org/abs/hep-ph/0611300}{arXiv:hep-ph/0611300}.

\bibitem{Beraudo2008}
 A.~Beraudo, J.-P.~Blaizot and C.~Ratti,
% \emph{Real and imaginary-time $Q\bar Q$ correlators in a thermal medium},
 Nucl. Phys. A \textbf{806} (2008) 312--338.
% \href{https://doi.org/10.1016/j.nuclphysa.2008.03.001}{doi:10.1016/j.nuclphysa.2008.03.001},
% \href{https://arxiv.org/abs/0712.4394}{arXiv:0712.4394}.


\bibitem{Burnier2015}
 Y.~Burnier, O.~Kaczmarek and A.~Rothkopf,
% \emph{Static quark-antiquark potential in the quark-gluon plasma from lattice QCD},
 Phys. Rev. Lett. \textbf{114} (2015) 082001.
% \href{https://doi.org/10.1103/PhysRevLett.114.082001}{doi:10.1103/PhysRevLett.114.082001}.

\bibitem{Bala2022}
 D.~Bala et al. (HotQCD Collaboration),
% \emph{Static quark-antiquark interactions at nonzero temperature from lattice QCD},
 Phys. Rev. D \textbf{105} (2022) 054513.
% \href{https://doi.org/10.1103/PhysRevD.105.054513}{doi:10.1103/PhysRevD.105.054513}.

\bibitem{Bazavov2024}
 A.~Bazavov et al. (HotQCD Collaboration),
% \emph{Unscreened forces in the quark-gluon plasma?},
 Phys. Rev. D \textbf{109} (2024) 074504.
%\href{https://doi.org/10.1103/PhysRevD.109.074504}{doi:10.1103/PhysRevD.109.074504}.


\bibitem{Ali:2025iux}
S.~Ali \textit{et al.} [HotQCD],
%``Thermal static potential and pseudoscalar quarkonium spectral functions from (2+1)-flavor lattice QCD,''
Phys. Rev. D \textbf{112} (2025) no.5, 054510.
%doi:10.1103/s8gw-n43f
%[arXiv:2505.11313 [hep-lat]].


\bibitem{Liu:2015ypa}
S.~Y.~F.~Liu and R.~Rapp,
%``An in-medium heavy-quark potential from the $Q\bar{Q}$ free energy,''
Nucl. Phys. A \textbf{941} (2015) 179-187.
%doi:10.1016/j.nuclphysa.2015.07.001
%[arXiv:1501.07892 [hep-ph]].
 
  
\bibitem{Brambilla2008}
N.~Brambilla, J.~Ghiglieri, P.~Petreczky and A.~Vairo,
%\emph{Static quark--antiquark pairs at finite temperature},
Phys. Rev. D \textbf{78} (2008) 014017.
%\href{https://doi.org/10.1103/PhysRevD.78.014017}{doi:10.1103/PhysRevD.78.014017},
%\href{https://arxiv.org/abs/0804.0993}{arXiv:0804.0993}.


\bibitem{Blaizot:2021xqa}
J.~P.~Blaizot and M.~{\'A}.~Escobedo,
%\emph{Phenomenological study of quarkonium suppression and the impact of the energy gap between singlets and octets}
Phys. Rev. D \textbf{104}, no.5 (2021) 054034.
%\href{https://doi:10.1103/PhysRevD.104.054034}.


 \bibitem{Carrington:2024ize}
M.~E.~Carrington, C.~Manuel and J.~Soto,
%\emph{High-Temperature QCD Static Potential beyond Leading Order},
Phys. Rev. Lett. \textbf{134}, no.1  (2025) 011905.
%\href{https://doi:10.1103/PhysRevLett.134.011905}{doi:10.1103/PhysRevLett.134.011905}.





\end{thebibliography}
\end{document}